\documentstyle[prl,aps,epsf]{revtex}

\begin{document}
\twocolumn[\hsize\textwidth\columnwidth\hsize\csname @twocolumnfalse\endcsname

\title{Hole Dynamics in Two-Dimensional Antiferromagnetic Mott Insulators}

\author{F.F. Assaad$^{1}$ and  M. Imada$^{2}$ \\
   $^{1}$ Institut f\"ur Theoretische Physik III, \\
   Universit\"at Stuttgart, Pfaffenwaldring 57, D-70550 Stuttgart, Germany. \\
   $^{2}$ Institute for Solid State Physics, University of Tokyo,  \\
   7-22-1 Roppongi,
   Minato-ku, Tokyo 106, Japan.  }
\maketitle

\begin{abstract}
The dispersion relation of a doped hole in the half-filled 
2D Hubbard model is shown to follow a $|\vec{k}|^4$ 
law around the  $(0, \pm \pi)$ and $( \pm \pi, 0 )$
points in the Brillouin zone. Upon addition of pair-hopping 
processes this dispersion relation is unstable 
towards a $|\vec{k}|^2$ law. The above follows from 
$T=0$ Quantum Monte calculations of the single particle 
spectral function $A(\vec{k}, \omega)$ on $16 \times 16$ lattices.  
We discuss finite dopings and argue that the added term restores 
coherence to charge dynamics and drives the system towards a 
$d_{x^2 - y^2}$ superconductor. 
\end{abstract}
]

The excitation spectrum  of a  band insulator is gapped in both
the spin and charge degrees of freedom.  In this case the single particle
Green function will essentially take a free particle form  since there
are no low lying excitations on which the added particle can scatter. 
The Mott insulator with long-range antiferromagnetic order has a charge
gap but no spin gap.  This leads to non-trivial hole dynamics which
will depend on the dynamical spin susceptibility as well as on the 
coupling between spin and charge degrees of freedom. Hole dynamics in
magnetic insulators have been addressed in the pioneering work
of Brinkman and Rice \cite{Brinkman70}. Progress in photoemission 
spectroscopy has provided us with an experimental realization of 
this problem \cite{Wells95}.
In this Letter,
we show numerically that hole dynamics in a two-dimensional Mott insulator
with long-range magnetic order may behave in radically different 
ways. The dispersion relation of a hole  doped into the half-filled
Hubbard model is very flat around the $(0,\pm \pi)$ and 
$(\pm \pi,0)$ points in the Brillouin zone. We will give numerical 
evidence that it follows  a $| \vec{k}|^4$
law. We argue that this flatness and resultant singular momentum  
dependence  of charge excitations provides us with a  microscopic basis
for understanding incoherent charge dynamics and unusual character of the 
metal-insulator transition.
Upon addition of a term  with matrix element $W$ which 
describes pair-hopping processes, this flat dispersion relation 
is unstable towards a $|\vec{k}|^{2}$ law. 
Although the added term for {\it small } values of $W$ does not alter 
the insulating state itself it restores coherence to charge dynamics
in the vicinity of the Mott transition.  

The model we consider reads:
\begin{equation}
\label{tU}
      H_{tUW} =  -\frac{t}{2} \sum_{\vec{i}} K_{\vec{i}} + 
          U \sum_{\vec{i}}
         \tilde{n}_{\vec{i},\uparrow}
         \tilde{n}_{\vec{i},\downarrow}
       - W  \sum_{\vec{i}} K_{\vec{i}}^{2} 
\end{equation}
where
\begin{equation}
     K_{\vec{i}} = \sum_{\sigma, \vec{\delta}}
   \left(c_{\vec{i},\sigma}^{\dagger} c_{\vec{i} + \vec{\delta},\sigma} +
     c_{\vec{i} + \vec{\delta},\sigma}^{\dagger} c_{\vec{i},\sigma} 
   \right).
\end{equation}
Here $\vec{\delta} = \pm \vec{a}_x, \pm \vec{a}_y $, 
$c_{\vec{i},\sigma}^{\dagger} $ creates an electron with $z$-component of
spin $\sigma$ on lattice site $\vec{i}$ and 
$\tilde{n}_{\vec{i},\sigma} 
= c_{\vec{i},\sigma}^{\dagger} c_{\vec{i},\sigma} -1/2$.
We consider a square lattice of linear size $L$ and impose periodic boundary
conditions in both lattice directions.  
The $t$-$U$-$W$ model  has 
been introduced and studied in Ref. 
\cite{Assaad96b,Assaad97,Assaad98,Assaad98a}. The results show that
at half-filling and finite values of $U/t$, 
$W$ triggers a quantum transition between the Mott
insulator and a $d_{x^2 - y^2}$  superconductor.   

To understand the  influence of the $W$-term on hole dynamics it is
convenient to consider the strong coupling limit where double occupancy
is prohibited. In this limit, the $W$-term
describes pair-hopping processes in the spin singlet,
$ -W \sum_{i,\vec{\delta},\vec{\delta'}}    \left(
\Delta^{\dagger}_{\vec{i}, \vec{\delta}} \Delta_{\vec{i}, \vec{\delta'}} 
+ \Delta_{\vec{i}, \vec{\delta}} \Delta^{\dagger}_{\vec{i}, \vec{\delta'}}
\right) $ 
and spin triplet 
$ W \sum_{i,\vec{\delta},\vec{\delta'},m}    \left(
T^{\dagger}_{\vec{i}, \vec{\delta},m} T_{\vec{i}, \vec{\delta'},m} 
+ T_{\vec{i}, \vec{\delta},m} T^{\dagger}_{\vec{i}, \vec{\delta'},m}
\right) $ 
channels. 
Here,  
$ \Delta^{\dagger}_{\vec{i}, \vec{\delta}} = \left(
c_{\vec{i},\uparrow}^{\dagger}
c_{\vec{i}+\vec{\delta},\downarrow}^{\dagger} -
c_{\vec{i}+\vec{\delta},\downarrow}^{\dagger}
c_{\vec{i},\uparrow}^{\dagger} \right)/\sqrt{2}$ and 
$T^{\dagger}_{\vec{i}, \vec{\delta},m}$ creates a spin triplet
on the  bond $\vec{i}$, $\vec{i} +\vec{\delta}$ with $z$-component of
spin $m$.
In the strong coupling limit, 
the states $T^{\dagger}_{\vec{i}, \vec{\delta},m}$ and 
$\Delta^{\dagger}_{\vec{i}, \vec{\delta}}$ form a complete basis for
a pair of electrons on the bond $\vec{i}$, $\vec{i} +\vec{\delta}$.
Thus, any spin configuration  surrounding a hole
may be written as a superposition of triplets and singlets on a set of 
bonds covering the lattice.  
One may now see how hole dynamics  are effected by the $W$-term:
the motion of a pair of electrons is nothing but the hopping  of a
hole within the same sublattice accompanied by a local rearrangement of 
spins. Due to zero point quantum fluctuations of the spin background, the
resultant state is not orthogonal to the ground state. 
Since the state in which the hole is doped has antiferromagnetic order,
we expect the {\it singlet}  pair-hopping processes to dominate the low-energy
physics.  Those processes  are similar to the three site term obtained
in a strong coupling expansion of the Hubbard model \cite{Hirsch85a}. 
The influence of those terms on hole dynamics has been addressed in  
Ref. \cite{Szczpanski90}.

At half-band filling auxiliary 
field quantum Monte Carlo (QMC) 
simulations provide an efficient tool for the study of the above model.
Partly due to the presence of particle-hole symmetry, the infamous
sign problem is avoidable. Our $T=0$ data are produced
with the projector QMC (PQMC) algorithm  \cite{Koonin86,Sorella89}
supplemented with a 
numerically stable method to compute imaginary time  displaced
correlation functions \cite{Assaad96a}. The imaginary time data is analytically 
continued to  the real axis with  the Maximum-Entropy method 
\cite{Linden95,Jarrell96}.  We 
have used a flat default model and taken into account  correlations  in 
imaginary time data with the use of the covariance matrix. 

\begin{figure}[ht]
\epsfxsize=10cm
\epsfxsize=7cm
\hfil\epsfbox{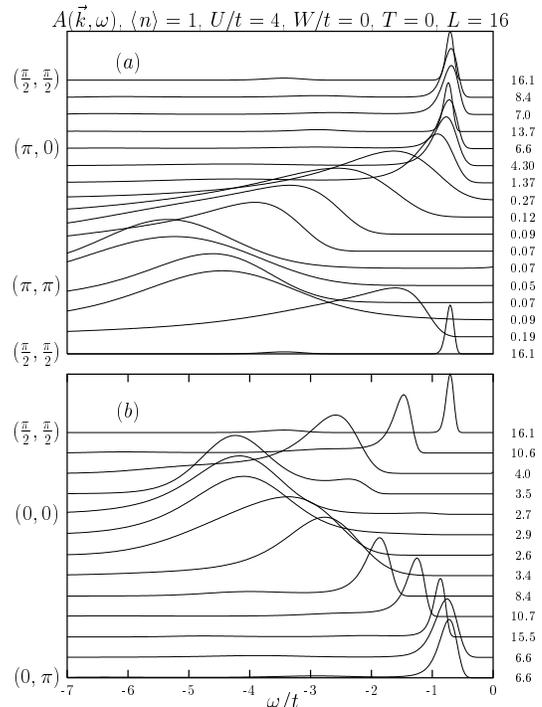}\hfil
\caption[]
{\noindent  $A(\vec{k}, \omega )$ at $T=0$  for the half-filled Hubbard
model at $U/t = 4$ on a $ 16 \times 16 $ lattice.
The considered path in the Brillouin zone is listed 
on the left hand side of the figure. We have normalized the raw data
by the factor listed on the right hand side of the figure.
This normalization sets the peak value of $A(\vec{k}, \omega )$ to unity
for all considered $\vec{k}$ vectors.
\label{tUAkom.fig} }
\end{figure}

\begin{figure}[ht]
\epsfxsize=10cm
\epsfxsize=7cm
\hfil\epsfbox{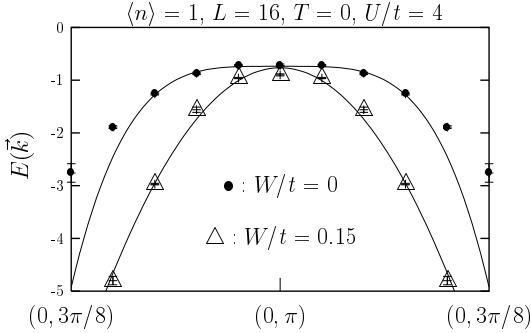}\hfil
\caption[]
{\noindent Dispersion relation, $E(\vec{k})$, as defined by the peak position
in $A(\vec{k},\omega)$.  The solid lines correspond to a least square fit
to the form $ a + b (\pi - k_y)^2 $  ($ a + b (\pi - k_y)^4 $)  for 
$W/t = 0$ ($W/t = 0.15$). For the fit we consider 
the $k_y$ range $[5 \pi/8,  \pi]$ for  both choices of $W/t$.
\label{dis01.fig} }
\end{figure}

We start by considering the half-filled Hubbard model at $U/t = 4$ and 
$T=0$. For this parameter set and after extrapolation
to the thermodynamic limit numerical simulations lead to the  conclusion
that the  ground state is an insulator with
single particle gap $\Delta_{qp} = 0.67 \pm 0.015$ \cite{Assaad96}
and long-range antiferromagnetic  order:
$m \equiv 
\lim_{L \rightarrow \infty}  
\sqrt{ 3 \langle m_z ((L/2,L/2) ) m_z( \vec{0} ) \rangle } 
= 0.39(5)$  \cite{Hirsch89,White89}.
Here, $m_z(\vec{i})  =  \tilde{n}_{\vec{i},\uparrow } - 
\tilde{n}_{\vec{i},\downarrow}$.
Fig. \ref{tUAkom.fig} plots 
$A(\vec{k},\omega) \equiv  {\rm Im } G(\vec{k}, \omega)$,
where $ G(\vec{k}, \omega)$ corresponds to the single-particle Green 
function. Due to particle-hole symmetry, 
$A(\vec{k},\omega) \equiv  A(\vec{k} + \vec{Q}, -\omega)$ with 
$\vec{Q} = (\pi, \pi)$ in units of the lattice constant. 
The sum rule: $ \int_{-\infty}^{0} A(\vec{k},\omega) {\rm d } \omega $
$ = \pi \sum_{\sigma}  $
$ \langle c_{\vec{k},\sigma}^{\dagger} c_{\vec{k},\sigma}  \rangle $
is satisfied.
We are primarily interested in the single-hole 
dispersion relation $E(\vec{k})$ as defined by the peak position in 
$A(\vec{k},\omega)$.  
Around  the $(0,\pi)$ and three equivalent points in the Brillouin zone, 
$E(\vec{k})$ shows a very flat structure.
Along the $(0,0)$ to $(0,\pi)$ direction $E(\vec{k})$ is compatible with 
a $|\vec{k}|^4$ law over an acceptable range
(see Fig. \ref{dis01.fig}). The same conclusion is reached when considering 
the $(\pi,\pi)$ to $(0,\pi)$ direction. 
Such a flat dispersion relation has been observed  numerically in 
$t$-$J$ model calculations \cite{Dagotto94a}. 
The overall flatness appears consistently with the flat 
dispersion observed in underdoped  cuprates \cite{Wells95,Gofron94,Shen95}.
The following points are of interest.
i) The energy difference 
$ \Delta E = E(\vec{k} = (\pi/2,\pi/2) ) - E(\vec{k} = (0,\pi) )$  is not
unambiguously distinguishable from zero within our resolution.  
$\Delta E$ = 
$0.045t$,  $0.06t$ and $0.015 t $ for the $L=8,12$ and $L=16$ lattices
respectively.  The uncertainty of our data, is in the same ball-park as
the above quoted values. 
We note that the flatness of the dispersion relation around the  $(0,\pi)$
point  should lead to a broader lineshape at $(0,\pi)$ than at 
$(\pi/2,\pi/2)$ \cite{Shen97}  thus leading to some ambiguity
in the definition of  $E(\vec{k})$.
In fact, defining
$E(\vec{k})$ as the leading edge rather than the peak position yields
a negative value for $\Delta E$ for the $L=16$ lattice. 
Lineshapes are notoriously hard to compute with the Maximum-Entropy method
and further large scale calculations are required to confirm this point. 
ii) A shadow-band due to the presence of long-range 
magnetic order is seen along the $(\pi/2,\pi/2)$ to
$(\pi,\pi)$ to $(\pi,0)$ direction.  Around the $(0,0)$ point 
a two {\it peak} feature is seen in the data.  A similar feature at larger 
couplings is seen in Ref. \cite{Preuss95}. 

\begin{figure}[ht]
\epsfxsize=10cm
\epsfxsize=7cm
\hfil\epsfbox{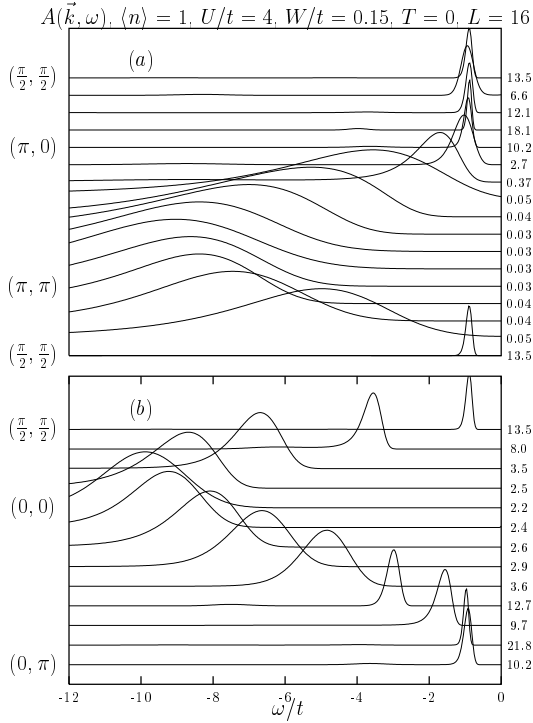}\hfil
\caption[]
{\noindent Same as Fig. \ref{tUAkom.fig} but  for $W/t = 0.15$. 
\label{tUWAkom.fig} }
\end{figure}

We now set $W/t  = 0.15$ and keep the other parameters constant,
$U/t = 4$, $T=0$, $\langle n \rangle = 1 $.  At this point in
parameter space, the ground state remains a Mott insulator with
$m = 0.24(1)$ and $ \Delta_{qp} = 0.54(6) $ \cite{Assaad98}. 
Fig. \ref{tUWAkom.fig}  plots 
$A(\vec{k},\omega)$ again on a $L=16$ lattice. The following
points are of importance. 
i) Upon inspection one notices
that the bandwidth is substantially enhanced by the inclusion of the
$W$-term. This high energy phenomena may be 
captured at a mean-field level  by the Ansatz:
$\langle n_{\vec{i}, \uparrow} \rangle + 
\langle n_{\vec{i}, \downarrow} \rangle = 1$,
$\langle n_{\vec{i}, \uparrow} \rangle  -
\langle n_{\vec{i}, \downarrow} \rangle = (-m_{HF})^{i_x + i_y}$
and $\langle K_{\vec{i}} \rangle  =  K$.
Self-consistency yields  the single particle gap 
$\Delta_{HF} \equiv Um_{HF}/2 =
0.43 t$ and a band ranging from   $-11.9 t$  to $11.9 t$.
The bandwidth agrees well with the Monte-Carlo data.  This approximation
{\it underestimates} the single particle gap thus showing  that is
does not capture the low-energy physics contained in the $W$-term. 
ii) Around the $(0,\pi)$ point, $E(\vec{k})$ is not
as flat as for the Hubbard model.
As shown in Fig. \ref{dis01.fig}, the data is compatible with a 
quadratic fit. To be more precise, 
as $W/t$ is enhanced the domain in
$\vec{k}$-space around the $(0, \pm \pi)$, $(\pm \pi,0)$ points 
which is compatible with a $|\vec{k}|^4$ fit is suppressed.
iii) As in Fig \ref{tUAkom.fig}  the shadow-band feature is present.
iv) The energy difference in the dispersion relation  between the
$ (\pi/2,\pi/2) $  and $(0,\pi)$ points  is not distinguishable
from zero.  \cite{Note2}

\begin{figure}[ht]
\epsfxsize=10cm
\epsfxsize=7cm
\hfil\epsfbox{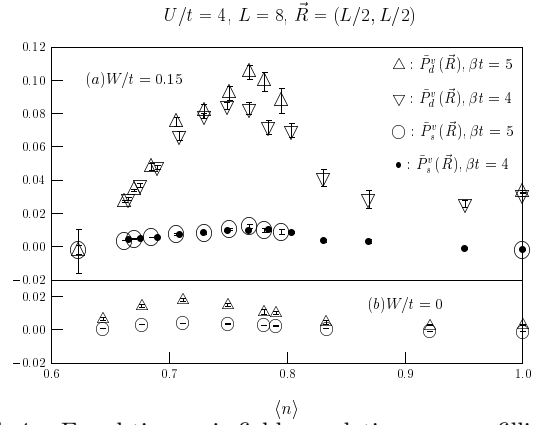}\hfil
\caption[]
{ \noindent
Equal-time pair field correlations versus filling in 
the $d$-wave and extended $s$-wave channels.  
$\tilde{P}_{d,s}^{v}(\vec{R})$ $\equiv$ 
$P_{d,s}^{v}(\vec{R})/$ $P_{d,s}^{v}(\vec{0})$.
The same scale is used for both (a) W/t = 0.15 and (b) W/t = 0.
\label{pair1.fig} }
\end{figure}

Next we study finite dopings.
Here, we are confronted to a sign problem so that the CPU time required 
to achieve a given precision scales exponentially with inverse 
temperature ($\beta$) and lattice size. The presented data is
produced with the finite temperature QMC method \cite{Hirsch85,White89}.
We first consider the vertex contribution to pairing 
correlations in the $d$-wave and extended $s$-wave channels. 
This quantity is defined by:
\begin{eqnarray}
\label{Pair_ver}
& &  P_{d,s}^{v} (\vec{r}) =  \langle \Delta_{d,s}^{\dagger}(\vec{r}) 
\Delta_{d,s}(\vec{0}) \rangle -
   \sum_{\sigma,\vec{\delta}, \vec{\delta}' }  f_{d,s}(\vec{\delta})
        f_{d,s}(\vec{\delta}')  \\
& & \left(
 \langle c^{\dagger}_{\vec{r},\sigma} c_{\vec{\delta}',\sigma}   \rangle
 \langle c^{\dagger}_{\vec{r}+\vec{\delta},-\sigma} c_{\vec{0},-\sigma}
  \rangle
+ \langle c^{\dagger}_{\vec{r},\sigma} c_{\vec{0},\sigma} \rangle
  \langle c^{\dagger}_{\vec{r}+\vec{\delta},-\sigma}
          c_{\vec{\delta}',-\sigma} \rangle \right)  \nonumber
\end{eqnarray}
where $ \Delta_{d,s}^{\dagger}(\vec{r})  =
\sum_{\sigma,\vec{\delta}}  f_{d,s}(\vec{\delta})
\sigma c^{\dagger}_{\vec{r},\sigma}
c^{\dagger}_{\vec{r} + \vec{\delta},-\sigma}$,
$f_{s}(\vec{\delta}) = 1$ and $f_{d}(\vec{\delta}) = 1 (-1)$
for $\vec{\delta} = \pm \vec{a_x}$ ($\pm \vec{a_y}$). Per definition, in
the absence of interaction $P_{d,s}^{v}$ vanishes  (i.e. in the 
above mentioned  mean field approximation which takes into account
band width effects, $P_{d,s}^{v}$ vanishes.)  Due to the strong variation
of $P_{d,s}^{v}(\vec{0})$ \cite{Note1} as a function of doping,
we consider the
quantity:  $\tilde{P}_{d,s}^{v}(\vec{r}) = 
P_{d,s}^{v}(\vec{r})/P_{d,s}^{v}(\vec{0})$ which measures the 
{\it decay} rate. This quantity is plotted versus doping 
in Fig. \ref{pair1.fig}. At the largest distance on our $8 \times 8$
lattice, the $W$-term substantially enhances 
the $d$-wave signal. We note that the same  conclusion is reached when
considering $P_{d,s} = \langle \Delta_{d,s}^{\dagger}(\vec{r})
\Delta_{d,s}(\vec{0}) \rangle $.

\begin{figure}[ht]
\epsfxsize=10cm
\epsfxsize=7cm
\hfil\epsfbox{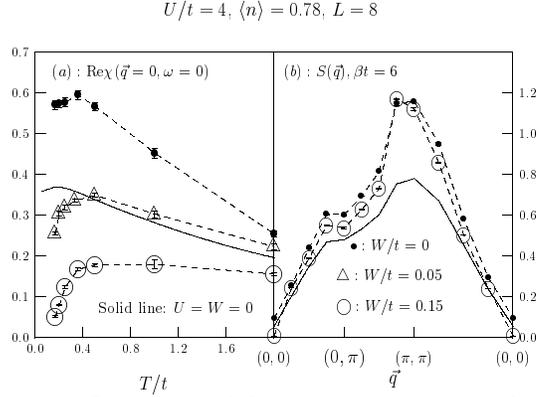}\hfil
\caption[]
{\noindent
(a) Spin susceptibility versus temperature for 
various values of $W/t$ and $\langle n \rangle = 0.78$.
(b) Spin structure factor  at $\beta t = 6$, $\langle n \rangle = 0.78$ 
for the Hubbard and $t$-$U$-$W$ models. 
In both (a) and (b) the solid line corresponds to the non-interacting
($U=W=0$) case. }
\label{spin.fig}
\end{figure}

The uniform spin susceptibility is plotted versus temperature in Fig. 
\ref{spin.fig}(a) at $\langle n \rangle = 0.78$, $U/t = 4$,
for various values of $W/t$.
Starting from the noninteracting case ($U$=$W$=$0$) and  turning on 
the Coulomb  repulsion to $U/t = 4$ enhances the spin susceptibility. 
At {\it high} temperatures, this enhancement may be understood 
within the random phase approximation. As $W/t$  grows, there is a
suppression of the spin susceptibility. There are two effects 
which cause this suppression. i) The enhancement of the bandwidth  as
a function of $W/t$.  ii)  The growth of $d$-wave pair correlations. 
Disentangling the contribution of those two effects at different
energy scales is non-trivial. The  spin structure factor 
$S(\vec{q}) \equiv \langle m_z (\vec{q}) m_z(-\vec{q}) \rangle$ 
is plotted in Fig. \ref{spin.fig} (b). The reduction of $S(\vec{q})$
at $\vec{q} = 0$  as $W/t$ is enhanced may be traced back to our 
discussion of the spin susceptibility since the latter quantity is
given by $\beta S(\vec{q}=0)$. In the vicinity of  
$\vec{q} = (\pi,\pi)$,  $S(\vec{q})$ shows a somewhat sharper feature at 
$W/t = 0.15$ than at $W/t = 0$, thus showing that the magnetic length
scale is {\it enhanced} by the $W$-term. 

In summary, we considered the $t$-$U$-$W$ model at $U/t = 4$ and
for two different choices of $W$: $W/t =0$, $W/t = 0.15$. In both cases,
the ground state at half-filling is a Mott insulator with long-range
antiferromagnetic order. However, the nature of hole dynamics in this 
antiferromagnetic background  is strongly  dependent on the choice of
$W/t$. 
In the case of the Hubbard model we concluded that the hole dispersion
relation is consistent with a $|\vec{k}|^4$ law around the  
$(0,\pm \pi)$ and $(\pm \pi,0)$ points in the Brillouin zone.  This
flat dispersion relation is compatible with the picture of incoherent
charge dynamics
and introduces a singular momentum dependence in the electron self-energy.
Recently, the authors of Ref. \cite{Tsunetsugu98,Imada98}
have computed the Drude weight on $t$-$J$ clusters and found results  consistent
$ D\sim \delta^2 $. The sum rule for the optical conductivity 
is proportional to the
doping $\delta$ in the case of the $t$-$J$ model. Thus, close to the
metal insulator transition, only a vanishingly small portion of 
the weight in the 
optical conductivity will be contained in the coherent Drude response. 
Introducing the $W$-term alters this situation. On one hand we have
shown here that the dispersion relation around the $(0,\pm \pi)$
and $(\pm \pi , 0)$ points follow a $|\vec{k}|^2$ law.  On the other hand, 
the Drude weight satisfies $D \sim \delta$ for the $t$-$J$-$W$ model
\cite{Tsunetsugu98,Imada98}. Thus, the $W$-term restores coherence
to charge dynamics in the vicinity of metal-insulator transition. 
However, the short-range antiferromagnetic spin correlations at finite doping
remain robust upon switching on $W$. 
In the doped phase, $d$-wave pairing correlations functions are  
substantially enhanced by the inclusion of $W$ thus lending support to 
the occurrence of a superconducting state. In terms of quantum phase
transitions the inclusion of the $W$-term alters the dynamical 
exponent from $z=4$ to $z=2$ 
\cite{Imada_rev,Assaad98,Tsunetsugu98,Imada98}.  More generally, 
the $W$-term exploits one of the many potential instabilities  
of the incoherent metallic state realized in the Hubbard model 
in the vicinity of the metal-insulator transition. As a more realistic
model for high-$T_c$ cuprates, smaller values of $W$ are to be considered 
so as to study the interplay between the pairing energy scale and the 
larger energy scale associated with the flat bands.

The numerical simulations were performed at 
the Supercomputer Center of the Institute for Solid State
Physics, University of Tokyo,  the HLRS-Stuttgart and HLRZ-J\"ulich. 
We thank M. Brunner, A. Muramatsu, H. Tsunetsugu and M. Zacher for 
conversations.

\end{document}